\documentclass[11pt]{article}
\usepackage{typearea}\typearea{12}
\usepackage{epsfig,amsmath,amsfonts,amssymb,mathtools}
\hypersetup{hidelinks}
\newtheorem{theorem}{Theorem}[section]
\newtheorem{lemma}[theorem]{Lemma}

\makeatletter
\@addtoreset{equation}{section}
\makeatother

\makeatletter
\long\def\@makecaption#1#2{{\small
\advance\leftskip1cm
\advance\rightskip1cm
\vskip\abovecaptionskip
\sbox\@tempboxa{#1: #2}%
\ifdim \wd\@tempboxa >\hsize
 #1: #2\par
\else
\global \@minipagefalse
\hb@xt@\hsize{\hfil\box\@tempboxa\hfil}%
\fi
\vskip\belowcaptionskip}}
\makeatother
\def\eq#1\en{\begin{equation}#1\end{equation}}  
\def\eqa#1\ena{\begin{align}#1\end{align}}
\def\eqg#1\eng{\begin{gather}#1\end{gather}}
\newcommand{\lb}[1]{\label{e:#1}}
\newcommand{\rlb}[1]{\eqref{e:#1}} 
\newcommand{\nl}{\notag\\}

\newcommand{\sumtwo}[2]%
{\mathop{\sum_{#1}}_{#2}}
\newcommand{\sumthree}[3]%
{\mathop{\mathop{\sum_{#1}}_{#2}}_{#3}}
\newcommand{\sumfour}[4]%
{\mathop{\mathop{\mathop{\sum_{#1}}_{#2}}_{#3}}_{#4}} 
\newcommand{\prodtwo}[2]%
{\mathop{\prod_{#1}}_{#2}}
\newcommand{\mintwo}[2]%
{\mathop{\min_{#1}}_{#2}}
\newcommand{\maxtwo}[2]%
{\mathop{\max_{#1}}_{#2}}
\newcommand{\maxthree}[3]%
{\mathop{\mathop{\max_{#1}}_{#2}}_{#3}}
\newcommand{\limtwo}[2]%
{\mathop{\lim_{#1}}_{#2}}
\newcommand{\suptwo}[2]%
{\mathop{\sup_{#1}}_{#2}}
\newcommand{\supthree}[3]%
{\mathop{\mathop{\sup_{#1}}_{#2}}_{#3}}
\newcommand{\supfour}[4]%
{\mathop{\mathop{\mathop{\sup_{#1}}_{#2}}_{#3}}_{#4}} 
\newcommand{\inftwo}[2]%
{\mathop{\inf_{#1}}_{#2}}
\newcommand{\infthree}[3]%
{\mathop{\mathop{\inf_{#1}}_{#2}}_{#3}}
\newcommand{\inffour}[4]%
{\mathop{\mathop{\mathop{\inf_{#1}}_{#2}}_{#3}}_{#4}} 

\newcommand\calP{{\cal P}}

\newcommand\calS{{\cal S}}

\newcommand{\sfA}{\mathsf{A}}
\newcommand{\sfB}{\mathsf{B}}
\newcommand{\sfC}{\mathsf{C}}
\newcommand{\sfD}{\mathsf{D}}
\newcommand{\sfE}{\mathsf{E}}

\newcommand{\bbC}{\mathbb{C}}

\newcommand{\bbR}{\mathbb{R}}

\newcommand{\qedm}{\rule{1.5mm}{3mm}}

\newcommand{\La}{\Lambda}

\newcommand{\hQ}{\hat{Q}}

\newcommand{\hA}{\hat{A}}
\newcommand{\hB}{\hat{B}}

\newcommand{\hH}{\hat{H}}
\newcommand{\hI}{\hat{I}}

\newcommand{\hW}{\hat{W}}
\newcommand{\hX}{\hat{X}}
\newcommand{\hY}{\hat{Y}}
\newcommand{\hZ}{\hat{Z}}

\newcommand{\sA}{\hat{\sfA}}
\newcommand{\sB}{\hat{\sfB}}
\newcommand{\sC}{\hat{\sfC}}
\newcommand{\sD}{\hat{\sfD}}
\newcommand{\sE}{\hat{\sfE}}

\newcommand{\la}{\lambda}

\newcommand{\ri}{\mathrm{i}}

\newcommand{\Supp}{\operatorname{Supp}}
\newcommand{\Diag}{\operatorname{Diag}}

\newcommand{\qA}{q_{\sA}}

\newcommand{\PL}{\calP_\La}

\newcommand{\Jx}{J_\mathrm{x}}
\newcommand{\Jy}{J_\mathrm{y}}

\newcommand{\hx}{h_\mathrm{x}}
\newcommand{\hy}{h_\mathrm{y}}
\newcommand{\hz}{h_\mathrm{z}}

\newcommand{\km}{\bar{k}}

\newcommand{\tq}{\tilde{q}}

\newcommand{\bunderline}[1]{\mkern2mu\underline{\mkern-2mu#1\mkern-1mu}\mkern1mu }
\newcommand{\ub}{\bunderline{u}}
\newcommand{\ut}{\bar{u}}
\newcommand{\ex}{e_\mathrm{x}}
\newcommand{\ey}{e_\mathrm{y}}
\newcommand{\ux}{u_\mathrm{x}}
\newcommand{\uy}{u_\mathrm{y}}
\newcommand{\uz}{u_\mathrm{z}}
\newcommand{\tfti}{\tfrac{1}{2\ri}}
\newcommand{\fti}{\frac{1}{2\ri}}
\newcommand{\msp}{\ \ \,}

\usepackage{color}
\definecolor{fluorescentpink}{rgb}{1.0, 0.08, 0.58}
\definecolor{forestgreen}{rgb}{0.13, 0.55, 0.13}

\begin{document}

\noindent
{\Large\bf 
Absence of nontrivial local conserved quantities in the quantum compass model on the square lattice}

\renewcommand{\thefootnote}{\fnsymbol{footnote}}
\medskip\noindent
Mahiro Futami and Hal Tasaki\footnote{%
Department of Physics, Gakushuin University, Mejiro, Toshima-ku, 
Tokyo 171-8588, Japan.
}
\renewcommand{\thefootnote}{\arabic{footnote}}
\setcounter{footnote}{0}

\begin{quotation}
\small\noindent
By extending the method developed by Shiraishi, we prove that the quantum compass model on the square lattice does not possess any local conserved quantities except for the Hamiltonian itself.
\end{quotation}

\tableofcontents

\section{Introduction}\label{S:introduction}
In 2019, Shiraishi \cite{Shiraishi2019} proved that the $S=\frac{1}{2}$ XYZ spin chain under a magnetic field does not possess any nontrivial local conserved quantities.
Since an integrable model generically has a series of nontrivial local conserved quantities \cite{JimboMiwa,Fadeev,Baxter,Takahashi,CauxMossel2011}, this result strongly suggests that the model is not integrable.
As far as we know, this was the first time that a specific quantum many-body model was proved to be non-integrable, or, more precisely, proved to exhibit a property that was never observed in integrable models.
See \cite{GrabowskiMathieu1995} for an earlier similar attempt.

Shiraishi's method was extended to the $S=\frac{1}{2}$ Ising chain under a slanted magnetic field \cite{Chiba2024a}, the PXP model \cite{ParkLee2024a}, and the $S=\frac{1}{2}$ spin chains with next-nearest-neighbor interactions \cite{Shiraishi2024} to show that these specific models do not possess nontrivial local conserved quantities.
The method was later applied to test for the presence/absence of nontrivial local conserved quantities exhaustively in general classes of spin chains, namely, the $S=1$ bilinear/biquadratic chains (with an anisotropy) \cite{ParkLee2024b,HokkyoYamaguchiChiba2024}, the $S=\frac{1}{2}$ chain with nearest neighbor symmetric interactions \cite{YamaguchiChibaShiraishi2024a,YamaguchiChibaShiraishi2024b}, and, finally,  the $S=\frac{1}{2}$ chain with nearest and next-nearest neighbor symmetric interactions \cite{Shiraishi2025}.
Remarkably, all models, except for those already known to be integrable, were found to have no nontrivial local conserved quantities.
Shiraishi's method was also extended to study the $S=\frac{1}{2}$ Ising, XY, and XYZ models on the hypercubic lattice in dimensions two or higher \cite{Chiba2024b,ShiraishiTasaki2024}.
It was concluded that all models except for the classical Ising model do not have nontrivial local conserved quantities.
Recently, Hokkyo developed an efficient, rigorous scheme for proving the absence of nontrivial local conserved quantities in a general class of quantum spin systems.  The scheme drastically simplifies some of the above-mentioned proofs of the absence of conserved quantities \cite{Hokkyo2025}.
See Discussion for how this new scheme can be applied to the problem treated in the present paper.

The same method of Shiraishi has also been useful in providing explicit characterizations of local conserved quantities in integrable models \cite{NozawaFukai2020,YamadaFukai2023,Fukai2023,Fukai2024}.

Let us also remark that, in the context of operator growth \cite{ParkerCaoAvdoshkinScaffidiAltman2019,NandyMatsoukasRoubeasMartinezAzconaDymarskyCampo2024}, Cao proved in 2021 that the $S=\frac{1}{2}$ Ising models under a magnetic field in one or higher dimensions exhibit a growth of moments that is characteristic to quantum chaotic models and is never observed in integrable models \cite{Cao2021}.
See Appendix~A.3 of \cite{ShiraishiTasaki2024} for a related discussion.

\medskip
In the present work, we extend Shiraishi's method to the quantum compass model on the square lattice (with or without a magnetic field) \cite{KugelKhomskii1982,NussinovvandenBrink2015,DoucotFeigel'manIoffeIoselevich2005,DorierBeccaMila2005,RichardsSorensen2024} and establish that the model does not possess any local conserved quantities other than the Hamiltonian itself.
Although the $S=\frac{1}{2}$ quantum Ising model and the XY/XYZ models, respectively, on the square lattice have already been studied by Chiba \cite{Chiba2024b} and Shiraishi and Tasaki \cite{ShiraishiTasaki2024}, respectively, the compass model does not fall into these standard classes of models.
We also stress that the compass model is not merely a model that has not been covered by previous studies.
It is an interesting test case regarding the presence or absence of nontrivial local conserved quantities, at least for the following three reasons:
First, the model is simple.
Note that the Hamiltonian \rlb{H} of the quantum compass model is merely a sum of the Ising models with $\hX\hX$ interactions in the horizontal direction and the Ising model with $\hY\hY$ interactions in the vertical direction.
Secondly, the compass model without a magnetic field is known to possess nontrivial global conserved quantities that are absent in the standard models such as the Ising, XY, or XYZ models \cite{DoucotFeigel'manIoffeIoselevich2005,DorierBeccaMila2005}.
See \rlb{GC} below.
Thirdly, its close cousin, namely, the quantum compass model on the hexagonal lattice, is nothing but the Kitaev honeycomb model \cite{Kitaev2006}.
The Kitaev model is famously known to possess many nontrivial local conserved quantities and is exactly solvable in certain senses \cite{Kitaev2006,BaskaranMandalShankar2007,YaoQi2010}.

While the above points suggest that the compass model on the square lattice is, in a sense, close to integrable models, we find that a proper extension of the proof in \cite{ShiraishiTasaki2024} shows that the only local conserved quantity of the model is the Hamiltonian.
We also find that the proof for the present model is simpler than any of the known proofs of the absence of nontrivial local conserved quantities \cite{Shiraishi2019,Chiba2024a,ParkLee2024a,ParkLee2024b,HokkyoYamaguchiChiba2024,YamaguchiChibaShiraishi2024a,YamaguchiChibaShiraishi2024b,Shiraishi2025,Chiba2024b,ShiraishiTasaki2024}.
The simplicity of the proof suggests that the model may serve as a test case for further studies of non-integrable models.

\section{Definitions and the main theorem}\label{S:main}
Let $\La=\{1,\ldots,L\}^2$ denote the $L\times L$ square lattice, whose site $u\in\La$ is sometimes denoted as $u=(\ux,\uy)$ with $\ux,\uy=1,\ldots,L$.
To define a $S=\frac{1}{2}$ spin system on $\La$, we  denote by $\hX_u$, $\hY_u$, and $\hZ_u$ the copies of the Pauli matrices at site $u\in\La$.
We study the quantum compass model without a magnetic field, whose Hamiltonian is
\eq
\hH=-\sum_{u\in\La}\bigl\{\Jx\,\hX_u\hX_{u+\ex}+\Jy\,\hY_u\hY_{u+\ey}\bigr\},
\lb{H}
\en
where $\ex=(1,0)$, $\ey=(0,1)$, $\Jx,\Jy\in\bbR$, $\Jx\ne0$, and $\Jy\ne0$ \cite{KugelKhomskii1982,NussinovvandenBrink2015,DoucotFeigel'manIoffeIoselevich2005,DorierBeccaMila2005}.
We impose the periodic boundary conditions.
The model with a magnetic field is discussed separately in section~\ref{S:mag}.

Although the model is simple, its exact solutions are not known.
It is easily verified that
\eqg
\hQ^{\rm Y}_{\uy}=\bigotimes_{\ux=1,\ldots,L}\hY_{(\ux,\uy)},\quad
\hQ^{\rm X}_{\ux}=\bigotimes_{\uy=1,\ldots,L}\hX_{(\ux,\uy)},
\lb{GC}
\eng
with any $\uy,\ux=1,\ldots,L$ commute with the Hamiltonian \cite{DoucotFeigel'manIoffeIoselevich2005,DorierBeccaMila2005}.

Let us now discuss the notion of local conserved quantities.
By a product of Pauli matrices (which we shall refer to as a product), we mean an operator of the form
\eq
\sA=\bigotimes_{u\in\Supp\sA}\hA_u,
\lb{A}
\en
with $\hA_u\in\{\hX_u,\hY_u,\hZ_u\}$, where $\Supp\sA\subset\La$, the support of $\sA$, is nonempty.
We denote the set of all products on $\La$ by $\PL$.
Note that the elements of $\PL$, with the identity $\hI$, span the whole space of operators of the spin system on $\La$.

In Shiraishi's proof and its extensions, it is crucial to define a proper measure of the ``size'' of a product.
While the width in one coordinate direction was a useful measure in the Ising, XY, and XYZ models on the square lattice \cite{Chiba2024b,ShiraishiTasaki2024}, we found that diagonal length is a proper measure in the quantum compass model.
The diagonal length of a product $\sA$, denoted as $\Diag\sA$, is defined as the minimum $k$ such that
\eq
0\le \ux+\uy-a\le k-1\ ({\rm mod}\ L),
\lb{0<a-a}
\en
for every $(\ux,\uy)\in\Supp\sA$ with some $a=1,\ldots,L$.
See Figure~\ref{f:shift}.
We say $\ut=(\ut_\mathrm{x},\ut_\mathrm{y})\in\Supp\sA$ is a top-right site of $\sA$ if $\ut_\mathrm{x}+\ut_\mathrm{y}-a=k-1$ with the same $a$ and $k$, and likewise, we say $\ub=(\ub_\mathrm{x},\ub_\mathrm{y})\in\Supp\sA$ is a bottom-left site of $\sA$ if $\ub_\mathrm{x}+\ub_\mathrm{y}-a=0$.

Fix a constant $\km$ such that $1\le \km\le L/2$, and write the candidate of a local conserved quantity with diagonal length $\km$ as
 \eq
 \hQ=\sumtwo{\sA\in\PL}{(\Diag\sA \le \km)}\qA\,\sA.
 \lb{Q}
 \en
The coefficients $\qA\in\bbC$ are arbitrary except that there exists at least one $\sA\in\PL$ such that $\qA\ne0$ and $\Diag\sA=\km$.
 We say that $\hQ$ is a local conserved quantity with diagonal length $\km$ if and only if
 \eq
 [\hH,\hQ]=0.
 \lb{HQ}
 \en
 Then, we prove the following.
 \begin{theorem}\label{T:main}
 The only local conserved quantities with diagonal length $\km$ such that $1\le\km\le L/2$ are constant multiples of the Hamiltonian $\hH$.
 \end{theorem}
   
\section{Proof}\label{S:proof}
\subsection{Basic strategy}\label{s:basic}
Our proof is based on the original strategy developed by Shiraishi \cite{Shiraishi2019,Shiraishi2024}, and is a variation of the proof for the $d$-dimensional XY and XYZ model by Shiraishi and Tasaki \cite{ShiraishiTasaki2024}.
In particular, our proof consists of two distinct steps, as in the first proof of Shiraishi \cite{Shiraishi2019}.
In this subsection, we prepare basic relations for the proof and treat the simplest case with diagonal length $\km=1$.
See Discussion for a strategy of the proof that makes use of Hokkyo's new scheme \cite{Hokkyo2025}.

For a product $\sA\in\PL$, we express its commutator with the Hamiltonian as a linear combination of products as
\eq
[\hH,\sA]=\sum_{\sB\in\PL}\la_{\sA,\sB}\,\sB.
\lb{HAexp}
\en
The coefficients $\la_{\sA,\sB}$ are readily determined by \rlb{H} and the commutation relations between the Pauli matrices.\footnote{
All that we need are the standard relations $\hX^2=\hY^2=\hZ^2=\hI$, $[\hX,\hY]=2\ri\hZ$, $[\hY,\hZ]=2\ri\hX$, $[\hZ,\hX]=2\ri\hY$, and $\{\hX,\hY\}=\{\hY,\hZ\}=\{\hZ,\hX\}=0$, where $\{\hA,\hB\}:=\hA\hB+\hB\hA$.
}
We say that $\sA$ generates $\sB$ if $\la_{\sA,\sB}\ne0$.
We then note that the commutator $[\hH,\hQ]$ for a general operator of the form \rlb{Q} is expressed as
\eq
[\hH,\hQ]=\sumtwo{\sA\in\PL}{(\Diag\sA \le \km)}\qA\,[\hH,\sA]
=\sum_{\sB\in\PL}
\Biggl(
\sumtwo{\sA\in\PL}{(\Diag\sA \le \km)}\la_{\sA,\sB}\,\qA
\Biggr)\,\sB,
\lb{HQexp}
\en
Since distinct $\sB$ are linearly independent, the condition \rlb{HQ} of a conserved quantity is equivalent to
\eq
\sumtwo{\sA\in\PL}{(\Diag\sA \le \km)}\la_{\sA,\sB}\,\qA=0,
\lb{relation}
\en
for any $\sB\in\PL$.
The coupled linear equations \rlb{relation} are our basic starting point.

As a trivial application of \rlb{relation}, we get the following Lemma.
It will be used repeatedly in the present paper.

\begin{lemma}\label{L:zero}
If there are $\sA,\sB\in\PL$ such that $\sA$ is the only product with $\Diag\le\km$ that generates $\sB$ (i.e., $\la_{\sA,\sB}\ne0$ and $\la_{\sA',\sB}=0$ for any $\sA'\in\PL\backslash\{\sA\}$ such that $\Diag\sA'\le\km$) then one has $\qA=0$.
\end{lemma}

For $3\le\km\le L/2$, we prove in sections~\ref{s:shift}, \ref{s:oddk}, and \ref{s:evenk} that the coupled equatios \rlb{relation} lead to the conclusion $\qA=0$ for all $\sA\in\PL$ with $\Diag\sA=\km$.
This contradicts the basic assumption about \rlb{Q} and hence implies that there are no local conserved quantities with diagonal length $\km$.

For $\km=2$, we show in section~\ref{s:k=2} that the only solutions of \rlb{relation} correspond to $\hQ=\eta\hH$ with a constant $\eta$.

Let us discuss the simplest case with $\km=1$.
Let $\sA$ be an arbitrary product with $\Diag\sA=1$ and take $u\in\Supp\sA$.
We then define
\eq
\sB=\begin{dcases}
-\tfti\,[\hY_{u}\hY_{u+\ey},\sA],&\text{if $\hA_{u}=\hX_{u}$};\\
\msp\tfti\,[\hX_{u}\hX_{u+\ex},\sA],&\text{if $\hA_{u}=\hY_{u}$};\\
-\tfti\,[\hX_{u}\hX_{u+\ex},\sA],&\text{if $\hA_{u}=\hZ_{u}$},
\end{dcases}
\en
where the prefactors are chosen so that $\sB\in\PL$.
Apparently, $\sA$ is the only product with diagonal length 1 that generates $\sB$.
We find $\qA=0$ from Lemma~\ref{L:zero}.
This proves the part of Theorem~\ref{T:main} with $\km=1$.

\subsection{First step: Shiraishi shift}\label{s:shift}
In the present subsection, we assume $2\le\km\le L/2$ and discuss the first step in the proof.
We shall show that the coefficient $\qA$ of a local conserved quantity $\hQ$ with diagonal length $\km$ is zero unless $\sA$ has a simple standard form.
An essential ingredient in the proof is a procedure that we call the Shiraishi shift.

Let us assume that there is a local conserved quantity \rlb{Q} with diagonal length $\km$.
We start by stating an important lemma that highlights a specific feature of the quantum compass model.
\begin{lemma}\label{L:basic}
Let $\sA\in\PL$ be such that $\Diag\sA=\km$ with $2\le\km\le L/2$.
One has $\qA=0$ unless the following two conditions are satisfied:
(1)~$\sA$ has a unique top-right site $\ut$ with (1a)~$\hA_{\ut}=\hX_{\ut}$, $\ut-\ex\in\Supp\sA$ or (1b)~$\hA_{\ut}=\hY_{\ut}$, $\ut-\ey\in\Supp\sA$.
(2)~$\sA$ has a unique bottom-left site $\ub$ with (2a)~$\hA_{\ub}=\hX_{\ub}$, $\ub+\ex\in\Supp\sA$ or (2b)~$\hA_{\ub}=\hY_{\ub}$, $\ub+\ey\in\Supp\sA$. 
\end{lemma}
{\em Proof:}\/
Let $\ut$ be one of (not necessarily unique) top-right sites of $\sA$, and let
\eq
\sB=\begin{dcases}
-\tfti\,[\hY_{\ut}\hY_{\ut+\ey},\sA],&\text{if $\hA_{\ut}=\hX_{\ut}$};\\
\msp\tfti\,[\hX_{\ut}\hX_{\ut+\ex},\sA],&\text{if $\hA_{\ut}=\hY_{\ut}$};\\
-\tfti\,[\hX_{\ut}\hX_{\ut+\ex},\sA],&\text{if $\hA_{\ut}=\hZ_{\ut}$},
\end{dcases}
\lb{Bdef}
\en
where the prefactors are chosen so that $\sB\in\PL$.
Note that the commutator appends the new site $\ut+\ey$ or $\ut+\ex$ to the support of $\sA$, and hence $\Diag\sB=\km+1$.
By definition $\sA$ generates $\sB$.
We then ask if there is another $\sA'$ with $\Diag\sA'\le\km$ that generates $\sB$.
If there is no such $\sA'$ then $\sA$ is the only product (with $\Diag\le\km$) that generates $\sB$, and we see $\qA=0$ from Lemma~\ref{L:zero}.

Note that any product  $\sA'$ with $\Diag\sA'\le\km$ that generates $\sB$ (other than $\sA$) must have $\Supp\sA'=\Supp\sB\backslash\{\ub\}$, where $\ub$ is the unique bottom-left site of $\sB$ (and hence the unique bottom-left site of $\sA$).
We also see that either $[\hX_{\ub}\hX_{\ub+\ex},\hA']$ or $[\hY_{\ub}\hY_{\ub+\ey},\hA']$ must be proportional to $\sB$.
This means (2a) or (2b) is valid.
We see condition (2) is necessary for nonzero $\qA$.

Repeating the same argument with top-right and bottom-left switched, we also find condition (1) is necessary for nonzero $\qA$.~\qedm

\medskip
The above proof contains the essential idea of the Shiraishi shift.

\begin{figure}
\centerline{\epsfig{file=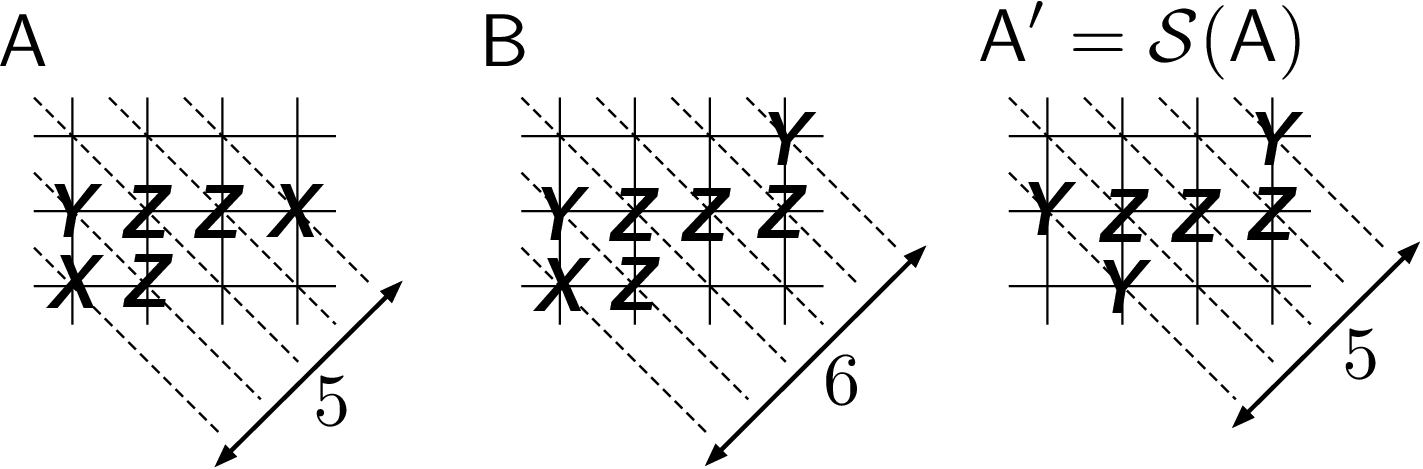,width=9truecm}}
\caption[dummy]{
An example of the Shiraishi shift in the case $\km=5$.
$\sA$ is a product with $\Diag\sA=5$ that satisfies conditions in Lemma~\ref{L:basic} with (1a) and (2a).
$\sB$ is defined by the commutator \rlb{SS1}. 
Since $\hY$ at the top-right is appended, the product has $\Diag\sB=6$.
We then truncated $\hX$ at the bottom-left by taking the commutator \rlb{SS2}.
The resulting product $\sA'$ with $\Diag\sA'=5$ is the Shiraishi shift $\calS(\sA)$.
Note that $\Supp\calS(\sA)$ has two bottom-left sites and does not satisfy condition (2).
This means $\calS^2(\sA)$ does not exist and also $q_{\calS(\sA)}=0$.
We also find $\qA=0$ from \rlb{q=q}.
}
\label{f:shift}
\end{figure}

To get the idea, let $\km\ge3$, and suppose that  $\sA\in\PL$ with $\Diag\sA=\km$ satisfies conditions (1) with (1a) and (2) with (2a).
See Figure~\ref{f:shift}.
We also assume $\hA_{\ub+\ex}=\hZ_{\ub+\ex}$.
As in \rlb{Bdef} we define
\eqa
\sB&=-\fti\,[\hY_{\ut}\hY_{\ut+\ey},\sA]
\nl&=-\fti\biggl(\bigotimes_{u\in\Supp\sA\backslash\{\ut\}}\hA_u\biggr)\otimes[\hY_{\ut},\hX_{\ut}]\otimes\hY_{\ut+\ey}
\nl&=\biggl(\bigotimes_{u\in\Supp\sA\backslash\{\ut\}}\hA_u\biggr)\otimes\hZ_{\ut}\otimes\hY_{\ut+\ey}\in\PL,
\lb{SS1}
\ena
where $\Diag\sB=\km+1$.
We next define
\eqa
\sA'&=-\fti\,[\hX_{\ub}\hX_{\ub+\ex},\sB]
\nl&=-\fti[\hX_{\ub}\hX_{\ub+\ex},\hX_{\ub}\hZ_{\ub+\ex}]\otimes\biggl(\bigotimes_{u\in\Supp\sB\backslash\{\ub,\ub+\ex\}}\hB_u\biggr)
\nl&=\hY_{\ub+\ex}\otimes\biggl(\bigotimes_{u\in\Supp\sB\backslash\{\ub,\ub+\ex\}}\hB_u\biggr),
\lb{SS2}
\ena
where we noted that $\hB_{\ub}=\hA_{\ub}=\hX_{\ub}$ and $\hB_{\ub+\ex}=\hA_{\ub+\ex}=\hZ_{\ub+\ex}$ because $\km\ge3$.
Observe that the commutator truncates the site $\ub$ from the support of $\sB$, and hence $\Diag\sA'=\km$.
An explicit calculation as in \rlb{SS1} shows
\eq
[\hX_{\ub}\hX_{\ub+\ex},\sA']=2\ri\,\sB.
\lb{SS3}
\en
Apparently $\sA$ and $\sA'$ are the only products with $\Diag\le\km$ that generate $\sB$.
The coefficients in the expansion \rlb{HAexp} are determined from \rlb{H}, \rlb{SS1} and \rlb{SS3} as $\la_{\sA,\sB}=-2\ri\Jy$ and $\la_{\sA',\sB}=2\ri\Jx$.
Then the basic condition \rlb{relation} for the coefficients reads
\eq
2\ri\Jy\,\qA-2\ri\Jx\,q_{\sA'}=0.
\lb{SS4}
\en
We write $\sA'$ as $\calS(\sA)$ and call it the Shiraishi shift of $\sA$.

We generalize this procedure and define the Shiraishi shift $\calS(\sA)\in\PL$ of $\sA\in\PL$  with $\Diag\sA=\km$.
We now consider general $\km$ with $2\le\km\le L/2$.
When $\sA$ does not satisfy conditions (1) or (2) in Lemma~\ref{L:basic}, we say that the Shiraishi shift $\calS(\sA)$ does not exist.
In this case, we see $\qA=0$ from Lemma~\ref{L:basic}.
 When $\sA$ satisfies (1) and (2), we define
\eq
\sB=
\begin{dcases}
-\tfti[\hY_{\ut}\hY_{\ut+\ey},\sA],&\text{when (1a) holds};\\
\msp\tfti[\hX_{\ut}\hX_{\ut+\ex},\sA],&\text{when (1b) holds},
\end{dcases}
\lb{SSdef1}
\en
and
\eq
\calS(\sA)=
\begin{dcases}
\pm\tfti[\hX_{\ub}\hX_{\ub+\ex},\sB],&\text{when (2a) holds};\\
\pm\tfti[\hY_{\ub}\hY_{\ub+\ey},\sB],&\text{when (2b) holds},
\end{dcases}
\lb{SSdef}
\en
provided that the commutator in \rlb{SSdef} is nonzero, where the sign should be chosen so that $\calS(\sA)\in\PL$.
We say that $\calS(\sA)$ does not exist if the commutator in \rlb{SSdef} is zero.
In this case, again, $\sA$ is the only product with $\Diag\le\km$ that generates $\sB$, and we see $\qA=0$ from Lemma~\ref{L:zero}.
When the shift $q_{\calS(\sA)}$ exsists, the coeffecients $\qA$ and $q_{\calS(\sA)}$ are related as in \rlb{SS4}.

Let us summarize the above observation as the following Lemma.
\begin{lemma}\label{L:shift}
Let $\sA\in\PL$ be such that $\Diag\sA=\km$ with $2\le\km\le L/2$.
When $\calS(\sA)$ does not exist one has $\qA=0$.
When $\calS(\sA)$ exists one has
\eq
q_{\calS(\sA)}=\la\,\qA,
\lb{q=q}
\en
with a nonzero constant $\la$.
\end{lemma}

By applying the Shiraishi shift multiple times, one can restrict products with possible nonzero coefficients to those in a standard form.
To describe the standard form, let us define a sequence of sites $u_1,u_2,\ldots\in\La$ by
\eq
u_j:=\begin{dcases}
u_1+\frac{j-1}{2}(\ex+\ey),&\text{for odd $j$};\\
u_1+\frac{j}{2}(\ex+\ey)-\ey,&\text{for even $j$},
\end{dcases}
\lb{uj}
\en
where $u_1$ is arbitrary.
Note that we employ the periodic boundary conditions.
We then define
\eq
\sC^k_1:=\hX_{u_1}\otimes\biggl(\bigotimes_{i=2}^{k-1}\hZ_{u_i}\biggr)\otimes\hY_{u_k},
\en
when $k$ is odd, and 
\eq
\sC^k_1:=\hX_{u_1}\otimes\biggl(\bigotimes_{i=2}^{k-1}\hZ_{u_i}\biggr)\otimes\hX_{u_k},
\lb{C1even}
\en
when $k$ is even.
We also define
\eq
\sC^k_j:=\calS^{j-1}(\sC^k_1),
\lb{Cjdef}
\en
for $j=2,3,\ldots$, where $\calS^{j-1}=\underbrace{\calS\circ\cdots\circ\calS}_{j-1}$.
See Figure~\ref{f:CDE}.
To be specific,
\eq
\sC^k_j=
\begin{dcases}
\hX_{u_j}\otimes\biggl(\bigotimes_{i=j+1}^{j+k-2}\hZ_{u_i}\biggr)\otimes\hY_{u_{j+k-1}},&\text{for odd $j$};\\
\hY_{u_j}\otimes\biggl(\bigotimes_{i=j+1}^{j+k-2}\hZ_{u_i}\biggr)\otimes\hX_{u_{j+k-1}},&\text{for even $j$},
\end{dcases}
\en
when $k$ is odd, and 
\eq
\sC^k_j=
\begin{dcases}
\hX_{u_j}\otimes\biggl(\bigotimes_{i=j+1}^{j+k-2}\hZ_{u_i}\biggr)\otimes\hX_{u_{j+k-1}},&\text{for odd $j$};\\
\hY_{u_j}\otimes\biggl(\bigotimes_{i=j+1}^{j+k-2}\hZ_{u_i}\biggr)\otimes\hY_{u_{j+k-1}},&\text{for even $j$},
\end{dcases}
\en
when $k$ is even.

Then, the following is the main result of the present section.
\begin{lemma}\label{L:onlyC}
Let $\sA\in\PL$ with $\Diag\sA=\km$ with $2\le\km\le L/2$.
Then one has $\qA=0$ unless $\sA=\sC^{\km}_1$ with $u_1=\ub$ or $\sA=\sC^{\km}_2$ with $u_2=\ub$.
\end{lemma}
\noindent{\em Proof:}\/
When $\km=2$, conditions (1) and (2) readily imply $\sA=\hX_{\ub}\hX_{\ut}$ with $\ut=\ub+\ex$ or $\sA=\hY_{\ub}\hY_{\ut}$ with $\ut=\ub+\ey$.
The former is $\sC^2_1$ with $u_1=\ub$, and the latter is $\sC^2_2$ with $u_2=\ub$.

We shall treat the case $3\le\km\le L/2$.

Suppose that $\sA$ satisfies condition (2) with (2a) and that $\calS(\sA)$ exists.
We shall examine necessary conditions that $\calS(\sA)$ satisfies (2).
Since $\hB_{\ub}=\hX_{\ub}$ by assumption, it is necessary that $\hB_{\ub+\ex}=\hY_{\ub+\ex}$ or $\hZ_{\ub+\ex}$ for the commutator in \rlb{SSdef} to be nonzero.
We claim that we must have $\hB_{\ub+\ex}=\hZ_{\ub+\ex}$ (which means $\hA_{\ub+\ex}=\hZ_{\ub+\ex}$).
To see this, note that $\ub+\ex$ is a bottom-left site of $\calS(\sA)$.
If $\hB_{\ub+\ex}=\hY_{\ub+\ex}$, \rlb{SSdef} implies $\calS(\sA)$ has $\hZ_{\ub+\ex}$ at its bottom-left site, which means that $\calS(\sA)$ never satifies (2a) or (2b).
With $\hB_{\ub+\ex}=\hZ_{\ub+\ex}$, we see $\calS(\sA)$ has $\hY_{\ub+\ex}$ at its bottom-left site.
This means $\calS(\sA)$ can only satisfy (2b).  We thus find $\ub+\ex+\ey\in\Supp\calS(\sA)$, which means  $\ub+\ex+\ey\in\Supp\sA$.
Finally, in order for $\ub+\ex$ to be the unique bottom-left site of $\calS(\sA)$, it must be that $\ub+\ex$ is the unique site in $\Supp\sA$ that satisfies $\ux+\uy=\ub_\mathrm{x}+\ub_\mathrm{y}+1$.
To sum up, we have seen that the bottom-left corner of $\sA$ is identical to that of $\sC^{\km}_1$ with $u_1=\ub$.

This argument can be repeated with conditions (2a) and (2b) used alternately.
After $\km-1$ repetitions, one finds, by also recalling (1), that $\sA$ is identical to $\sC^{\km}_1$ with $u_1=\ub$.

When $\sA$ satisfies condition (2) with (2b), the same argument (with (2a) and (2b) switched) shows that $\sA$ is identical to $\sC^{\km}_2$ with $u_2=\ub$.~\qedm

\subsection{Second step for odd $\km$}\label{s:oddk}
In the second step of the proof, we show that $q_{\sC^{\km}_j}=0$ for $\km$ with $3\le\km\le L/2$.
Then we see from Lemma~\ref{L:onlyC} that $\qA=0$ for all $\sA\in\PL$ with $\Diag\sA=\km$.
This proves the main part of Theorem~\ref{T:main}.

\begin{figure}
\centerline{\epsfig{file=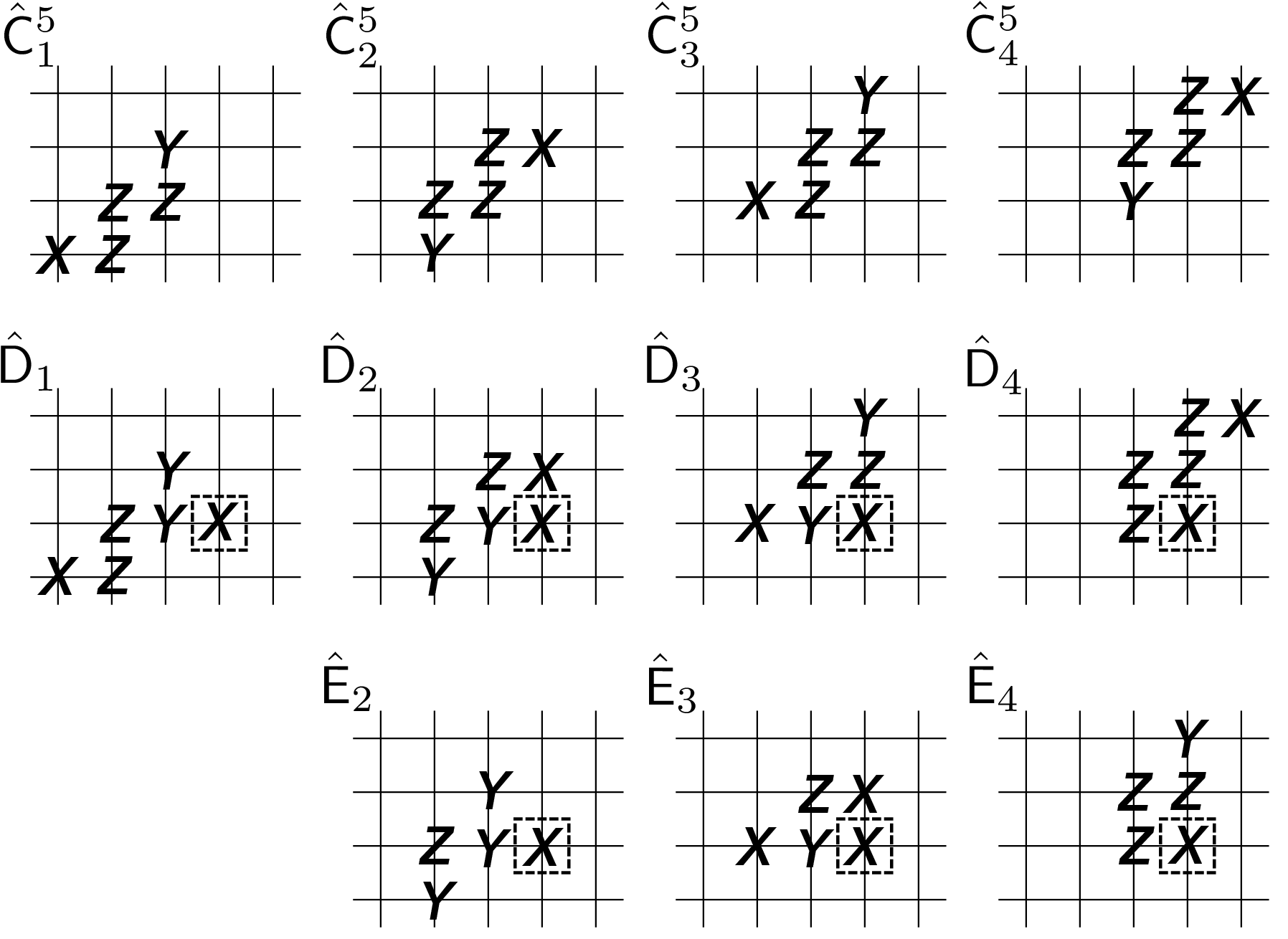,width=11truecm}}
\caption[dummy]{
Products $\sC^{\km}_j$, $\sD_j$, and $\sE_j$ for $\km=5$.
The dotted squares indicate $\hX_{u_{\km-1}+\ex}$ that are ``appended'' by the commutators in \rlb{Dj} and \rlb{Ej}.
}
\label{f:CDE}
\end{figure}

In this section, we treat the case with odd $\km$.
We first recall that $\sC^{\km}_j$ and $\sC^{\km}_{j+1}$ are the only products with $\Diag\le\km$ that generate $\sC^{\km+1}_j$.
(Note that $\sC^{\km+1}_j$ plays the role of $\sB$ in section~\ref{s:shift}.)
Examining the coefficients as we did in \rlb{SS4}, we find that the basic relation \rlb{relation} reads
\eqg
2\ri\Jx\,q_{\sC^{\km}_j}+2\ri\Jx\,q_{\sC^{\km}_{j+1}}=0,\quad\text{for odd $j$};\\
-2\ri\Jy\,q_{\sC^{\km}_j}-2\ri\Jy\,q_{\sC^{\km}_{j+1}}=0,\quad\text{for even $j$}.
\eng
By writing $q=q_{\sC^{\km}_1}$, we then have
\eq
q_{\sC^{\km}_j}=
\begin{cases}
q,&\text{for odd $j$};\\
-q,&\text{for even $j$}.
\end{cases}
\lb{qjodd}
\en

Let us now define
\eq
\sD_j:=
\begin{dcases}
-\tfti\,[\hX_{u_{\km-1}}\hX_{u_{\km-1}+\ex},\sC^{\km}_j],&j=1,\ldots,\km-2;\\
\msp\tfti\,[\hX_{u_{\km-1}}\hX_{u_{\km-1}+\ex},\sC^{\km}_{\km-1}],&j=\km-1.
\end{dcases}
\lb{Dj}
\en
The signs are chosen so that $\sD_j\in\PL$, by noting that the Pauli operator at site $u_{\km-1}$ in $\sC^{\km}_j$ is $\hZ$ if $j=1,\ldots,\km-2$ and is $\hY$ if $j=\km-1$.
Note that $\Diag\sD_j=\km$.
We also define
\eq
\sE_j:=
\begin{dcases}
-\tfti\,[\hX_{u_{\km-1}}\hX_{u_{\km-1}+\ex},\sC^{\km-1}_j],&j=2,\ldots,\km-2;\\
\msp\tfti\,[\hX_{u_{\km-1}}\hX_{u_{\km-1}+\ex},\sC^{\km-1}_{\km-1}],&j=\km-1.
\end{dcases}
\lb{Ej}
\en
We see $\sE_j\in\PL$ and $\Diag\sE_j=\km-1$.
See Figure~\ref{f:CDE}.

We shall examine the basic relation \rlb{relation} by setting $\sB=\sD_j$ for $j=1,\ldots,\km-1$.
To this end, we observe that, for $j=2,\ldots,\km-2$, the product $\sD_j$ is generated by $\sC^{\km}_j$, $\sE_j$, $\sE_{j+1}$, and other products with diagonal length $\km$.
Since Lemma~\ref{L:onlyC} guarantees that all these unnamed products have zero coefficients, the corresponding relations \rlb{relation} read
\eqg
2\ri\Jx\,q_{\sC^{\km}_j}-2\ri\Jx\,q_{\sE_j}+2\ri\Jy\,q_{\sE_{j+1}}=0,\quad \text{for even $j\in\{2,\ldots,\km-3\}$},
\lb{r1}\\
2\ri\Jx\,q_{\sC^{\km}_j}+2\ri\Jy\,q_{\sE_j}-2\ri\Jx\,q_{\sE_{j+1}}=0,\quad \text{for odd $j\in\{2,\ldots,\km-3\}$},
\lb{r2}
\eng
and
\eq
2\ri\Jx\,q_{\sC^{\km}_{\km-2}}+2\ri\Jy\,q_{\sE_{\km-2}}+2\ri\Jx\,q_{\sE_{\km-1}}=0.
\lb{r3}
\en
Likewise $\sD_1$ is generated by $\sC^{\km}_1$, $\sE_2$, and other irrelevant products, and $\sD_{\km-1}$ is generated by $\sC^{\km}_{\km-1}$, $\sE_{\km-1}$, and other irrelevant products.
The corresponding relations \rlb{relation} are
\eqg
2\ri\Jx\,q_{\sC^{\km}_1}-2\ri\Jx\,q_{\sE_2}=0,
\lb{r4}\\
-2\ri\Jx\,q_{\sC^{\km}_{\km-1}}-2\ri\Jx\,q_{\sE_{\km-1}}=0.
\lb{r5}
\eng

By using \rlb{qjodd}, writing $\tq_j=q_{\sE_j}$, and introducing the anisotropy parameter $\kappa=\Jx/\Jy\ne0$, the relations \rlb{r1}, \rlb{r2}, \rlb{r3}, \rlb{r4}, and \rlb{r5} are rewritten as
\eqg
q-\tq_2=0,\\
q+\tq_j-\kappa\,\tq_{j+1}=0,\quad \text{for $j=2,4,\ldots,\km-3$},\\
q+\kappa\,\tq_j-\tq_{j+1}=0,\quad \text{for $j=3,5,,\ldots,\km-4$},\\
q+\kappa\,\tq_{\km-2}+\tq_{\km-1}=0,\\
q-\tq_{\km-1}=0.
\eng
Summing up all the equations, one gets $(\km-1)q=0$, which leads to $q=0$ and hence $q_{\sC^{\km}_j}=0$.

\subsection{Second step for even $\km$}\label{s:evenk}
Let us discuss the case with even $\km$ such that $4\le\km\le L/2$.
We shall only write down main definitions and equations, as the argument is completely parallel to that in section~\ref{s:oddk}.

Corresponding to \rlb{qjodd}, we now have
\eq
q_{\sC^{\km}_j}=
\begin{cases}
q,&\text{for odd $j$};\\
\kappa\,q,&\text{for even $j$},
\end{cases}
\lb{qjeven}
\en
with $\kappa=\Jx/\Jy\ne0$.
We again define
\eq
\sD_j:=
\begin{dcases}
\msp\tfti\,[\hY_{u_{\km-1}}\hY_{u_{\km-1}+\ey},\sC^{\km}_j],&j=1,\ldots,\km-2;\\
-\tfti\,[\hY_{u_{\km-1}}\hY_{u_{\km-1}+\ey},\sC^{\km}_{\km-1}],&j=\km-1,
\end{dcases}
\en
and
\eq
\sE_j:=
\begin{dcases}
\msp\tfti\,[\hY_{u_{\km-1}}\hY_{u_{\km-1}+\ey},\sC^{\km-1}_j],&j=2,\ldots,\km-2;\\
-\tfti\,[\hY_{u_{\km-1}}\hY_{u_{\km-1}+\ex},\sC^{\km-1}_{\km-1}],&j=\km-1,
\end{dcases}
\en
where $\sD_j,\sE_j\in\PL$, $\Diag\sD_j=\km$, and $\Diag\sE_j=\km-1$.
The basic relation \rlb{relation} with $\sB=\sD_j$ for $j=1,\ldots,\km-1$ become
\eqg
-2\ri\Jy\,q_{\sC^{\km}_1}-2\ri\Jx\,q_{\sE_2}=0,\\
-2\ri\Jy\,q_{\sC^{\km}_j}+2\ri\Jy\,q_{\sE_j}+2\ri\Jy\,q_{\sE_{j+1}}=0,\quad \text{for $j=2,4,\ldots,\km-4$},\\
-2\ri\Jy\,q_{\sC^{\km}_j}-2\ri\Jx\,q_{\sE_j}-2\ri\Jx\,q_{\sE_{j+1}}=0,\quad \text{for $j=3,5,\ldots,\km-3$},\\
-2\ri\Jy\,q_{\sC^{\km}_{\km-2}}+2\ri\Jy\,q_{\sE_{\km-2}}-2\ri\Jy\,q_{\sE_{\km-1}}=0,\\
2\ri\Jy\,q_{\sC^{\km}_{\km-1}}-2\ri\Jx\,q_{\sE_{\km-1}}=0.
\eng
They again lead to $(\km-1)q=0$, from which we conclude $q_{\sC^{\km}_j}=0$.

\subsection{The case $\km=2$}\label{s:k=2}
The case with $\km=2$ is indeed easy at this stage.
Let $\hQ$ be any local conserved quantity with $\km=2$ and express it as \rlb{Q}.
We already know from Lemma~\ref{L:onlyC} that $\qA\ne0$ only when $\sA=\hX_u\hX_{u+\ex}$ or $\sA=\hY_u\hY_{u+\ey}$.
The only remaining task is to show that the coefficients $\qA=\eta\Jx$ for the former and $\qA=\eta\Jy$ for the latter, where $\eta$ is an arbitrary nonzero constant.
But this is already shown in \rlb{qjeven} for the products that can be Shiraishi shifted with each other.
Note that (only for the case with $\km=2$) we can apply the Shirashi shift from bottom-right to top-left to the same products.
This shows
\eq
q_{\hX_u\hX_{u+\ex}}=\eta\Jx,\quad q_{\hY_u\hY_{u+\ey}}=\eta\Jy,
\en
with a common $\eta$ for all $u\in\La$.
This completes the proof of Theorem~\ref{T:main}.

\section{The model with a magnetic field}\label{S:mag}
The Hamiltonian of the quantum compass model under a magnetic field $(\hx,\hy,\hz)\in\bbR^3$ is
\eq
\hH'=-\sum_{u\in\La}\bigl\{\Jx\,\hX_u\hX_{u+\ex}+\Jy\,\hY_u\hY_{u+\ey}+\hx\,\hX_u+\hy\,\hY_u+\hz\,\hZ_u\bigr\}.
\lb{Hh}
\en
It was observed recently in \cite{RichardsSorensen2024} that for the special choice $\hx=2\Jx$, $\hy=2\Jy$, and $\hz=0$, \rlb{Hh} is rewritten as
\eq
\hH'=-\sum_{u\in\La}\bigl\{\Jx(\hX_u+1)(\hX_{u+\ex}+1)+\Jy(\hY_u+1)(\hY_{u+\ey}+1)\bigr\}+(\Jx+\Jy)L^2.
\lb{Hh2}
\en
For $\Jx<0$ and $\Jy<0$, the Hamiltonian \rlb{Hh2} with even $L$ is frustration-free, i.e., all the terms in\rlb{Hh2} can be minimized simultaneously to yield exact ground states of the model.
See \cite{RichardsSorensen2024} for details.

Here, we claim that for any $\hx$, $\hy$, and $\hz$, including those values corresponding to the frustration-free models, Theorem~\ref{T:main} is still valid as it is provided that $\Jx\ne0$ and $\Jy\ne0$.

The proof is not difficult.
Suppose that a product $\sB$ is generated from $\sA$ with a magnetic field term in the Hamiltonian \rlb{Hh}, i.e., $\sB=\pm\fti[\hW_u,\sA]\in\PL$, where $\hW_u=\hX_u$, $\hY_u$, or $\hZ_u$.
It is crucial to note that one has $\Supp\sB=\Supp\sA$ and hence $\Diag\sB=\Diag\sA$ in this case.

In section~\ref{s:shift}, we considered $\sB\in\PL$ with $\Diag\sB=\km+1$ and examined products with $\Diag\le\km$ that generate $\sB$.
This means we do not have to consider any product that generates $\sB$ with a magnetic field term.
We see that all the conclusions in section~\ref{s:shift}, including Lemma~\ref{L:onlyC},  are valid.

In sections~\ref{s:oddk} and \ref{s:evenk}, we examined products that generate $\sD_j$.
Although there are several new products with $\Diag=\km$ that generate $\sD$ with magnetic filed terms, they are certainly not of the standard form and are known to have zero coefficients from Lemma~\ref{L:onlyC}.
Again, we see that the conclusions of these sections remain valid.
This completes the proof of the absence of local conserved quantities with $3\le\km\le L/2$.

Finally, the two-body part of a conserved quantity with $\km=2$ is determined by the argument in section~\ref{s:k=2}.
It is obvious that the one-body part should be proportional to the magnetic field part in the Hamiltonian \rlb{Hh} since the argument at the end of section~\ref{s:basic} is valid and shows there are no conserved quantities with $\km=1$.

\section{Discussion}
We proved that the $S=\frac{1}{2}$ quantum compass model on the square lattice (with or without a magnetic field) has no local conserved quantities other than constant multiples of the Hamiltonian.
The result strongly suggests that the model is non-integrable.
This is in sharp contrast with the compass model on the hexagonal lattice, which is nothing but the Kitaev honeycomb model \cite{Kitaev2006,BaskaranMandalShankar2007,YaoQi2010}.

Our poof is a variation of that of Shiraishi and Tasaki \cite{ShiraishiTasaki2024}, which is based on the original strategy developed by Shiraishi in \cite{Shiraishi2019}.
Thanks to the simplicity of the model, our proof is simpler than any known proofs of the absence of nontrivial local conserved quantities \cite{Shiraishi2019,Chiba2024a,ParkLee2024a,ParkLee2024b,HokkyoYamaguchiChiba2024,YamaguchiChibaShiraishi2024a,YamaguchiChibaShiraishi2024b,Shiraishi2025,Chiba2024b,ShiraishiTasaki2024}.
We hope that the quantum compass model on the square lattice will serve as a simple test case for further analysis of non-integrable quantum spin models.

After the present work was completed, we learned from Akihiro Hokkyo that his general scheme developed in \cite{Hokkyo2025} could be applied to the quantum compass model.
Let us discuss the proof according to this strategy.
Hokkyo's scheme essentially relies on the condition of the Hamiltonian called injectivity.
See (4) in \cite{Hokkyo2025}.
Although the Hamiltonian \rlb{H} of the compass model apparently seems to violate the injectivity, it satisfies the condition after a proper (and indeed natural) identification of the corresponding one-dimensional model.
For $x=1,\ldots,L$, let us regard the collection of $L$ spins on sites $u=(\ux,\uy)$ with $\ux+\uy=x$ (mod $L$) as a single (huge) ``spin."
Then, the compass model is rewritten as a periodic chain of $L$ ``spins" with an injective Hamiltonian.
Then Hokkyo's result shows that there are no $\km$-local conserved quantities with $3\le\km\le L/2$ provided that a 3-local quantity $\hQ$ such that $[\hH,\hQ]$ is at most 2-local does not exist.
This means that we only need to control candidates of local conserved quantities with diagonal length $\km=3$ to prove the main result about the absence of local conserved quantities.
We can, therefore, consider only the cases $\km\le3$ in section~\ref{s:shift} and only the case with $\km=3$ in section~\ref{s:oddk}.
Although our proof, which is fully explained in the present paper, is straightforward and simple, it is true that the application of Hokkyo's scheme drastically simplifies the proof.

We believe that essentially the same method, with the aid of Hokkyo's scheme \cite{Hokkyo2025}, applies to the $S=\frac{1}{2}$ quantum compass model on the cubic lattice with the Hamiltonian
\eq
\hH=\sum_{u\in\{1,\ldots,L\}^3}\{\hX_u\hX_{u+\ex}+\hY_u\hY_{u+\uy}+\hZ_u\hZ_{u+\uz}\},
\en
to show that it possesses no conserved quantities other than the Hamiltonian.
The model is very likely to be non-integrable in spite of its apparent similarity with the Kitaev honeycomb model \cite{Kitaev2006}.

 \bigskip
\noindent{\small
{\em Acknowledgement:}\/ 
We thank Akihiro Hokkyo for useful discussions and for allowing us to include in Discussion his observation on the applicability of his scheme to the compass model.
We also thank Hosho Katsura, Naoto Shiraishi, Masafumi Udagawa, and Kanji Yamada for their valuable discussions.
H.T. was supported by JSPS Grants-in-Aid for Scientific Research No. 22K03474.}


\end{document}